\documentstyle{article}

\newtheorem{theorem}{THEOREM}
\newtheorem{lemma}{LEMMA}

\begin{document}

\begin{center} {\Large\bf  On the local Hamiltonian structure of vector
fields \\}
{\small Dedicated to Ansgar Schnitzer who died tragically in a \\
motorcycle accident on October 16th  1993\\}
\vspace{5mm}
{\large \bf P. Crehan \\
Dept. of Mathematics, University of Kyoto\\
PAT@KUSM.KYOTO-U.AC.JP} \end{center}

\begin{abstract} We derive a canonical form for smooth vector fields on
$\Re^{n+1}$. We use this to demonstrate the local multi-Hamiltonian nature of
the corresponding flows. Associated with the canonical form is an
inhomogenious linear PDE whose solutions provide conserved measures.
These can be used to construct the local Hamiltonians. \end{abstract}

\section{Introduction}

The study of Poisson manifolds has received some attention recently as
many interesting Poisson structures arise in the context of integrable
systems, solvable models and quantum groups \cite{nlps}, \cite{marmo1},
\cite{marmo2}. For this reason it is important to understand the scope
of the Poisson manifold framework. Formally it is very similar to that
of symplectic geometry, but extends it in a simple and non-trivial way.

Without the machinery of symplectic geometry and Hamiltonian mechanics,
the concepts of solvability and full integrability, although still
meaningful are slightly less clear \cite{ham_nonham}. Many dynamical systems
of interest in mathematical physics are not usually treated by
Hamiltonian methods. Examples include the damped harmonic oscillator,
Lotka-Volterra systems and the Lorenz flow. Furthermore it has been
claimed that if a system is not Hamiltonian (Lagrangian), then it cannot
be consistently quantised \cite{nolag_noquant}. Our results on the local
structure of vector fields suggest however that the property of being
Hamiltonian has broader scope than is generally afforded.

Wintner pointed out that every smooth vector field is locally fully
integrable \cite{wintner}. More sophisticated versions of this
straigtening-out theorem arise as a consequence of the Lie-Frobenius
theory of distributions of vector fields \cite{westenholz}. The
straightening-out theorem on $\Re^{2n}$ has an important consequence for
autonomous Hamiltonian vector fields. The $2n-1$ local integrals must
close as an algebra under the Poisson bracket. If they did not they
would generate a further independent integral and the flow lines would
consist only of single points. In this way we know that autonomous
Hamiltonian vector fields are locally fully integrable. The existence of
integrals is intimately connected with the existence of symmetries.
Doebner et al. clarified this connection at the local level when they
showed that the local symmetries of a Hamiltonian system depend only on
the dimension of the space \cite{doebner}. We extend these local
analyses by proving some theorems, one of which asserts that every
smooth vector field locally has a multiple Hamiltonian structure. A
vector field is considered to be Hamiltonian with respect to a Poisson
structure rather than a symplectic structure. This is true for odd as
well as even dimensional manifolds. The odd dimensional case is not to
be confused with a local contact structure.

In this article we deal with the Hamiltonian structure of generic
vector fields only at the local level. Nevertheless we consider
our results motivation for a study of general dynamical systems from a Poisson
manifold point of view. Perhaps within a framework of the geometry of
Poisson manifolds we can reach a more unified understanding of
integrability. Perhaps the study of deformations of Poisson structures
can provide a theoretical basis for considering the quantisation of a
more general class of flows. Explicit examples which illustrate our
results for simple but interesting flows on the plane will be published
elsewhere.

\section{The local structure of vector fields} We consider flows on
$\Re^{n+1}$ generated by vector fields $v=v^i\partial_i$ according to

\begin{eqnarray} D_t &=& v^i \partial_i\label{theflow} \end{eqnarray}

\noindent where $v^i$, $i=0,\ldots,n \in C^1(\Re^{n+1})$. From the
theory of Lie and Frobenius, in the neighbourhood of a non-critical
point of the flow there exist co-ordinate systems
$\{x,\Phi^{(1)},\ldots,\Phi^{(n)}\}$ in which $v=f \partial_x$ for some
smooth function $f$. Furthermore the one dimensional distribution $V$
generated by $v$, is tangent to the foliation of $\Re^{n+1}$ by the one
dimensional submanifolds of the intersection of the level surfaces
$\Phi^{(1)}=c_1$, $\ldots$, $\Phi^{(n)}=c_{n}.$ In particular this means
that $\partial_x \Phi^{(k)}=0$ for $k=1,\ldots,n.$ In the original co-
ordinate system we have

	$$v^i \partial_i \Phi^{(1)} =0,\ldots,v^i
\partial_i\Phi^{(n)}=0.$$

\noindent The components of $v$ can therefore be represented as an
antisymmetric product of gradient fields. We are led directly to the
following lemma for smooth vector fields.

\begin{lemma}\label{lemma:canonicalform} In a sufficiently small but
finite neighbourhood $U$ of a non-critical point of a vector field
$v=v^i\ \partial_i$ on $\Re^{n+1}$, there exist functions $\rho\in
C^1(U)$ and $\Phi^{(1)},\ldots,\Phi^{(n)}\in C^2(U)$ such that

\begin{eqnarray} v^i&=&\rho^{-1}\ \epsilon^{i j_1 \ldots j_{n}}\
\partial_{j_1}\Phi^{(1)}\ldots \partial_{j_n}\Phi^{(n)}
\label{canonicalform} \end{eqnarray}

\noindent where $\epsilon^{i j_1\ldots j_n}$ is the totally
antisymmetric tensor. \end{lemma}

\noindent The inverse of $\rho$ has been used instead of $\rho$ for
convenience later on. Before we state the main theorem we need to know a
little bit about Poisson structures of rank 2. Poisson structures on a
manifold $\cal M$ are tensor fields $\Lambda^{ij}$ which satisfy the
following tensorial relations

\begin{eqnarray} \Lambda^{ij}+\Lambda^{ji}&=&0,
\label{c1}\\ \Lambda^{it}\partial_t\Lambda^{jk}+
\Lambda^{jt}\partial_t\Lambda^{ki}+
\Lambda^{kt}\partial_t\Lambda^{ij}&=&0.            \label{c2}
\end{eqnarray}

\noindent Tensors which satisfy (\ref{c1}) and (\ref{c2}) can be used to
construct Poisson brackets. The Poisson bracket of two functions $A, B
\in C^\infty(\cal M)$ being given by $\{A, B\ \} = \Lambda^{ij}\
\partial_i A\ \partial_j B$. It turns out that all brackets which
satisfy the axioms for a Poisson bracket are of this form. A review of
the theory of Poisson manifolds with references can be found in
\cite{weinstein}.

In $\Re^2$ the most general Poisson structure is of the form

\begin{eqnarray} \Lambda^{ij}&=& \Theta\ \epsilon^{ij}
\label{psr2}\end{eqnarray}

\noindent where $\Theta\in C^1(\Re^2)$ and $\epsilon^{ij}$ is the
totally antisymmetric tensor. In $\Re^3$ a Poisson tensor must have rank
2. In general it has the form

\begin{eqnarray} \Lambda^{ij}&=&\Theta\ \epsilon^{ijk}\ \partial_k \Phi
\label{psr3}\end{eqnarray}

\noindent where $\Theta\in C^1(\Re^3)$, $\Phi\in C^2{\Re^3}$ and
$\epsilon^{ijk}$ is the totally antisymmetric tensor on $\Re^3$. This
was shown in \cite{crehan} but has more recenly appeared in a slightly
different form in \cite{marmo1}. Due to the antisymmetry condition
eqn(\ref{c1}) $\Lambda$ must have even rank. In $\Re^4$ Poisson tensors
must therefore have rank 2 or 4. Those of rank 2 are of the general form
\cite{crehan}

\begin{eqnarray} \Lambda^{ij}&=&\Theta\ \epsilon^{ijkl}\ \partial_k
\Phi^{(1)} \ \partial_l \Phi^{(2)}. \label{psr4}\end{eqnarray}

\noindent A glance at equations (\ref{psr2}), (\ref{psr3}) and
(\ref{psr4}) suggests the following lemma.

\begin{lemma}\label{lemma:r2ps} Tensors on $\Re^{m+2}$ of the form

$$\Lambda^{i j}=\Theta\ \epsilon^{i j s_1\ldots s_m}\
\partial_{s_1}\Phi^{(1)}\ldots \partial_{s_n}\Phi^{(n)}$$

\noindent where $\Theta\in C^1(\Re^{m+2})$, $\Phi^{(k)} \in
C^2(\Re^{m+2})$ $k=1,\ldots,m$ and $\epsilon^{ijs_1 \ldots s_n}$ is the
totally antisymmetric tensor, provide Poisson structures of rank 2.
\end{lemma}

\noindent Eqn (\ref{c1}) is automatically satisfied. To verify eqn
(\ref{c2}) it is best to move to the co-ordinate system
$\{X,Y,\Phi^{(1)},\ldots,\Phi^{(m)}\}$. In these co-ordinates the only
non-zero terms of $\Lambda$ are $\Lambda^{XY}=-\Lambda^{YX}$. $\Lambda$
is therefore of rank 2 and eqn (\ref{c2}) can be verified by a trivial
calculation. The converse is also true and can be proven by a similar
argument.

The expression for $v$ given by eqn (\ref{canonicalform}) can be
rewritten as

	$$ v^i = \Lambda^{i j}_{(k)}\ \partial_{j} \Phi^{(k)} $$

\noindent where

$$ \Lambda^{i j}_{(k)}=\rho^{-1}\ \epsilon^{i s_1\ldots s_{k-1} j
s_{k+1} \ldots s_{n}}\ \partial_{s_1}\Phi^{(1)}\ldots \partial_{s_{k-
1}}\Phi^{(k-1)}\
\partial_{s_{k+1}}\Phi^{(k+1)}\ldots\partial_{s_n}\Phi^{(n)}.$$

\noindent According to lemma (\ref{lemma:r2ps}) each of these
$\Lambda_{(k)}$ is a Poisson tensor of rank 2. They are independent in
the sense that they determine different foliations of $\Re^{n+1}$ by 2
dimensional symplectic leaves. In this sense eqn(\ref{theflow}) can be
thought of as generating a flow which is Hamiltonian with Hamiltonian
$\Phi^{(k)}$ with respect to the Poisson structure $\Lambda_{(k)}$. We
can even drop the need for labels on the $\Phi^{(k)}$ by using instead
of a $\Phi^{(k)},$ a single
$\Phi=\Phi^{(1)}+\Phi^{(2)}+\ldots+\Phi^{(n)}$. Now we can state the
main theorem

\begin{theorem}\label{theorem:mhs} In a finite but sufficiently small
neighbourhood $U$ of a non-critical point of a smooth vector field
$v=v^i\partial_i$ on $\Re^{n+1}$, there exists a function $\Phi\in
C^2(U)$ and a family of $n$ smooth independent rank 2 Poisson structures
$\Lambda_{(k)}$ $k=1,\ldots,n$ so that

$$ v^i = \Lambda^{ij}_{(k)}\ \partial_j\Phi\hspace{1truecm}
k=1,\ldots,n.$$ \end{theorem}

In this sense all smooth vector fields locally have a multiple
Hamiltonian structure. The trajectories of the flow (\ref{theflow}) are
determined by the $\Phi^{(k)}$ in eqn(\ref{canonicalform}). They are not
affected by $\rho$, although the spreading of classical wavepackets
along the trajectories is. The $\Phi^{(k)}$ are readily interpreted as
local Hamiltonians but to understand the geometrical and physical
significance of $\rho$ we have to do a little more work.

\section{An equation for $\rho$, conserved measures and the local
Hamiltonians}

The functional form of the $\Phi^{(k)}$ is not unique and nor is that of
$\rho$. As long as $F$ is a well behaved non-trivial function of
$\Re^{n}$, eqn(\ref{canonicalform}) is invariant under the replacements

\begin{eqnarray} \Phi^{(k)}&\to&
F(\Phi^{(1)},\ldots,\Phi^{(n)})\label{phisub}\\ \rho&\to&\rho\
\partial_k F(\Phi^{(1)},\ldots,\Phi^{(n)}). \nonumber \end{eqnarray}

\noindent One can show by direct substsitution of
eqn(\ref{canonicalform}) that $\rho$ satisfies the following first order
linear PDE

\begin{eqnarray} \rho (\nabla\cdot v) + v\cdot(\nabla \rho)
=0.\label{therhoeqn} \end{eqnarray}

\noindent It is customary to study such equations as initial value
problems. One takes a set of initial data $\Gamma$ on a hypersurface
$\Sigma$ in $\Re^{n+1}$. If $\Sigma$ is nowhere tangent to $v$, then the
initial data can be extended from $\Sigma$ to $\Re^{n+1}$ so that
eqn(\ref{therhoeqn}) is satisfied. In the theory of partial differential
equations, the surfaces of intersection of the level surfaces of the
local Hamiltonians $\Phi^{(k)}$ are known as the characteristics. In
other words $\Sigma$ must be tranverse to each of the level surfaces of
the local Hamiltonians. It is easy to show that if $\rho$ is a solution
of eqn(\ref{therhoeqn}), then so is $\rho
F(\Phi^{(1)},\ldots,\Phi^{(n)})$ where $F\in C^1( \Re^n)$.

One can also show that the ratio of two independant solutions of
eqn(\ref{therhoeqn}) provides a conservation law. To see this compute
$D_t(\rho_1 \rho_2^{-1})$ using $D_t = v\cdot \nabla$ and
eqn(\ref{therhoeqn}). More precisely we can state the following lemma

\begin{lemma}\label{lemma:ratioofrho} Given $\Sigma$ a hypersurface in
$\Re^{n+1}$ nowhere tangent to the vector field $v$. If $\rho_A$ and
$\rho_B$ are solutions of $\rho (\nabla\cdot v) + v\ \cdot(\nabla
\rho)=0$ with inital data $\Gamma_A$ and $\Gamma_B$ on $\Sigma$. Then
$\rho_A \rho^{-1}_B$ is an integral of the flow $\dot{x}^i=v^i.$
\end{lemma}

A theorem due to Liouville tells us that the volume element is conserved
by flows which are Hamiltonian with respect to the standard symplectic
structure. We know that this is not true for general dynamical systems.
To see how we might characterize measures conserved by (\ref{theflow}),
we consider $\Delta$ - a smooth function on $\Re^{n+1}$ with compact
support. $\Delta_t$ is defined by transporting $\Delta$ with the flow
for a time $t$. A measure on $\Re^{n+1}$ is denoted $d\mu=\mu\
dx^0\wedge\ldots\wedge dx^{n}=\mu\  dV.$ The measure of $\Delta$ is
denoted by $\mu(\Delta)$ and given by $\mu(\Delta) = \int \Delta\ d\mu.$
To determine for what $\rho$ is $D_t \mu(\Delta_t)=0$ we compute

$$D_t \mu(\Delta_t) = \int (D_t \Delta_t) d\mu = -\int \rho \Delta_t\
D_t(\rho^{-1}\ \mu)\  dV. $$

\noindent On the support of $\rho$ the conserved measure must therefore
satisfy $D_t(\rho^{-1} \mu)=0$. One solution is given by $\mu=\rho$. By
inspection any $\mu = \rho F(\Phi^{(1)},\ldots,\Phi^{(n)})$ also
provides a solution. From lemma(\ref{lemma:ratioofrho}) if $\rho$
satisfies eqn(\ref{therhoeqn}) then so must $\mu = \rho
F(\Phi^{(1)},\ldots,\Phi^{(n)})$. Furthermore if $\rho^{\prime}$ is
any other solution of eqn(\ref{therhoeqn}),
lemma(\ref{lemma:ratioofrho}) implies that $D_t ( \rho
/\rho^{\prime})=0$. We therefore conclude with the following theorem.

\begin{theorem}\label{theorem:localmandi} Measures of the form $d\mu =
\rho\ dx^0\wedge\ldots\wedge dx^{n}$ are preserved by the flow
$\dot{x}^i = v^i$ if and only if $v\cdot(\nabla \rho)+\rho (\nabla\cdot
v)=0.$ Ratios of independent local measures provide local integrals.
\end{theorem}

\noindent The $\rho$ which appeared in the canonical form
eqn(\ref{canonicalform}) is therefore a local conserved measure.

\section{Discussion} Eqn(\ref{canonicalform}) provides an expression for
flows in  terms of local Hamiltonians $\Phi^{(k)}$ and locally conserved
measures $\rho$. Theorem(\ref{theorem:localmandi}) links these through a
linear PDE (\ref{therhoeqn}) derived from the flow. In general the
$\Phi^{(k)}$ are single valued only locally. They cannot be extended to
globally defined single valued functions on $\Re^{n+1}$. Even in the
case of fully integrable systems on $\Re^{2m}$, which are Hamiltonian
with respect to the standard symplectic structure, it is rare for all of
the integrals to be single valued. A Hamiltonian system on $\Re^{2m}$
is said to be fully integrable if there exist $m$ globally defined
single valued functions including the Hamiltonian, which are in
involution under the Poisson bracket. If we call these $\Phi^{(k)}$ for
$k=1,\ldots,m,$ the method of solution is to find a further set of
functions $\phi_{(k)}$, which are conjugate to the $\Phi^{(k)}$. In the
co-ordinate system
$(\Phi^{(1)},\ldots,\Phi^{(n)},\phi^{(1)},\ldots,\phi^{(n)})$ the
equations of motion are $D_t \phi^{(k)} = -\partial_{\Phi^{(k)}}
H(\Phi),$ and $ D_t \Phi^{(k)} = 0,$ for $k=1,\ldots,m.$ In these
coordinates the equations are linear and the system can be solved.
Although only $m$ integrals have been used in the solution of the
problem, it is possible to construct a further $m-1$ integrals of the
motion $\Phi^{(i\ i+1)}$ as follows

\begin{eqnarray*} \Phi^{(i\ i+1)}&=&\phi^{(i)} D_t \Phi^{(i+1)} -
\phi^{(i+1)} D_t \Phi^{(i)}. \end{eqnarray*}

\noindent The functions $\Phi^{(i)}$ for $i=1,\ldots,m$ and $\Phi^{(i\
i+1)}$ for $i=1,\ldots,m-1$ constitute the full set of $2m-1$ local
integrals. The simplest illustration of this is of course the system of
two uncoupled harmonic oscillators. The Hamiltonian of this system with
respect to the standard symplectic structure on $\Re^4$ is $(p_x^2+w_x^2
x^2)/2+(p_y^2+w_y^2 y^2)/2.$ In accordance with
lemma(\ref{lemma:canonicalform}) we can write the equations of motion in
the form

\begin{eqnarray*} \dot{x}^i &=& \rho^{-1}\ \epsilon^{i j k l}\
\partial_j \Phi^{(1)}\ \partial_{k} \Phi^{(2)}\ \partial_l \Phi^{(12)},
\end{eqnarray*}

where $\rho=1,$ and the three local Hamiltonians are

\begin{eqnarray*} \Phi^{(1)}&=&{1\over 2}(p_x^2+w^2_x x^2),\\
\Phi^{(2)}&=&{1\over 2}(p_y^2+w^2_y y^2),\\ \Phi^{(12)}&=&{1\over
w_x}\arctan(w_x {x\over p_x})- {1\over w_y}\arctan(w_y {y\over p_y}).
\end{eqnarray*}

\noindent In general $\Phi^{(12)}$ is multivalued and unless the winding
frequencies are commensureable, it takes infinitely many values. This
gives rise to a flow which is ergodic on the surface defined by the
intersection of the level surfaces of $\Phi^{(1)}$ and $\Phi^{(2)}$. The
case where all of the integrals are single valued is considered to be
degenerate. In the case of a generic vector field, there are no single
valued integrals. A generic vector field can therefore be thought of as
a flow, none of whose Hamiltonians are isolating. A careful study of the
relationship between constants of motion and degeneration for systems
which are Hamiltonian with respect to standard Hamiltonian structures is
provided by Onofri and Pauri \cite{degeneracy}. It seems reasonable to
extend this classification to arbitrary vector fields on the basis of
lemma(\ref{lemma:canonicalform}) and theorem(\ref{theorem:mhs}).

A lot of work has been done on the subject of integrability.  Much of
which centers around applications of Painlev\'e analysis. According to
one reviewer \cite{ham_nonham}, the conceptual basis for this is less
clear in the non-Hamiltonian case than in the Hamiltonian case.
Painlev\'e analysis proceeds by complexifying the dynamical system. It
then deals with the integrability of the complex system by investigating
the branching of solutions of the system in the complex time plane.
Rigorous results such as those of Ziglin \cite{ziglin} provide criteria
for integrability in terms of the monodromony group of the system in the
complex time plane. If a system is fully integrable when it is
complexified, then it must be integrable. On the other hand Kozlov gives
an example of a system which although non-integrable in the complex
domain is completely integrable in the real domain \cite{kozlov}. The
concept of integrability of a real dynamical system is therefore
slightly different from that which is directly addressed by Painlev\'e
analysis.

Theorem(\ref{theorem:mhs}) suggests an alternative framework for
investigating arbitrary smooth vector fields. A framework which
incorporates essential geometrical ideas of Hamiltonian mechanics. It
might provide an interesting standpoint from which to understand
concepts of integrability. In particular it might be instructive to
consider the singularities of solutions of the associated PDE
(\ref{therhoeqn}) and the relationship they bear to those encountered in
Painlev\'e analysis.

It seems natural to pursue a deformation approach to the quantisation of
vector fields \cite{deformation}, based on the geometry of Poisson
manifolds rather than that of symplectic manifolds. Barriers to the
quantisation of systems indicated in \cite{nolag_noquant} could be
removed at least locally in this way. An interesting question is whether
the symmetry between local Hamiltonians which is present at the
classical level, will also be present at the quantum level.

\section{Acknowledgement} I would like to thank Michio Jimbo for his
hospitality at Kyoto University Dept. of Mathematics, as well as for
his interest in and discussion of this work. I would also like to thank
Hiroshi Kokubu for his interest and discussion.

 \end{document}